\documentclass{sig-alternate}

\begin{document}
%

\newfont{\mycrnotice}{ptmr8t at 7pt}
\newfont{\myconfname}{ptmri8t at 7pt}
\let\crnotice\mycrnotice%
\let\confname\myconfname%

\permission{Authors' pre-submission draft.}
\conferenceinfo{}{{\mycrnotice{}}}
\copyrightetc{}
\crdata{}

\clubpenalty=10000 
\widowpenalty = 10000

\title{Detection Bank: An Object Detection Based Video Representation for Multimedia Event Recognition}
%
%
%
%
%

\numberofauthors{3} 
%
\author{
%
%
\alignauthor
Tim Althoff\\
       \affaddr{UC Berkeley EECS/ICSI}\\
       \email{althoff@icsi.berkeley.edu}
\alignauthor
Hyun Oh Song\\
       \affaddr{UC Berkeley EECS/ICSI}\\
       \email{song@eecs.berkeley.edu}
\alignauthor 
Trevor Darrell\\
       \affaddr{UC Berkeley EECS/ICSI}\\
       \email{trevor@eecs.berkeley.edu}
}

\maketitle
\begin{abstract}
While low-level image features have proven to be effective representations for visual recognition tasks such as object recognition and scene classification, they are inadequate to capture complex semantic meaning required to solve high-level visual tasks such as multimedia event detection and recognition. Recognition or retrieval of events and activities can be improved if specific discriminative objects are detected in a video sequence. In this paper, we propose an image representation, called \textit{Detection Bank}, based on the detection images from a large number of windowed object detectors where an image is represented by different statistics derived from these detections. This representation is extended to video by aggregating the key frame level image representations through mean and max pooling. We empirically show that it captures complementary information to state-of-the-art representations such as Spatial Pyramid Matching and Object Bank. These descriptors combined with our Detection Bank representation significantly outperforms any of the representations alone on TRECVID MED 2011 data.
\end{abstract}




\section{Introduction}
 
\begin{figure*}[ht]
\centering
{\includegraphics[width=0.40\textwidth]{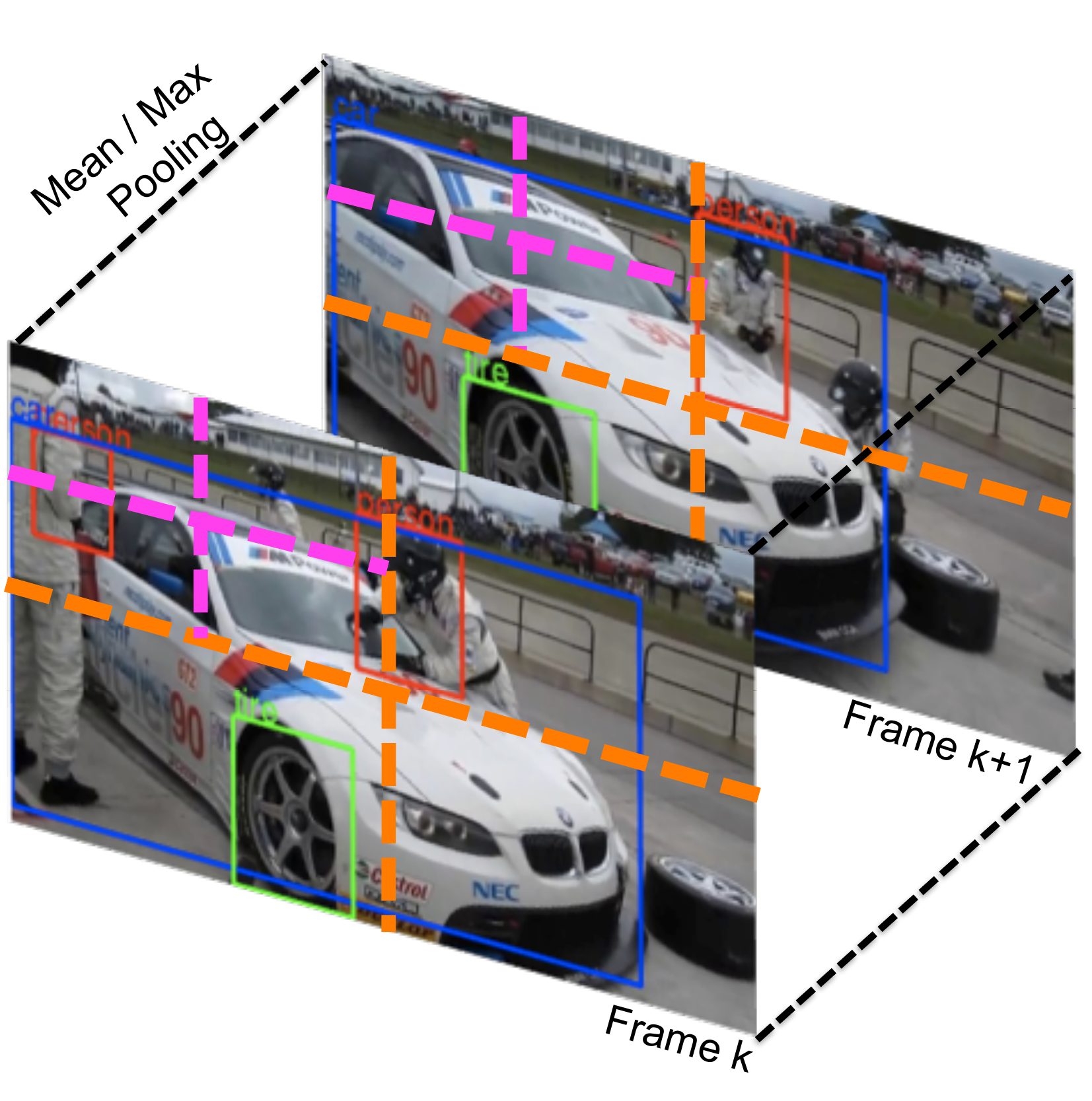}} \hspace{10pt}
{\includegraphics[width=0.35\textwidth]{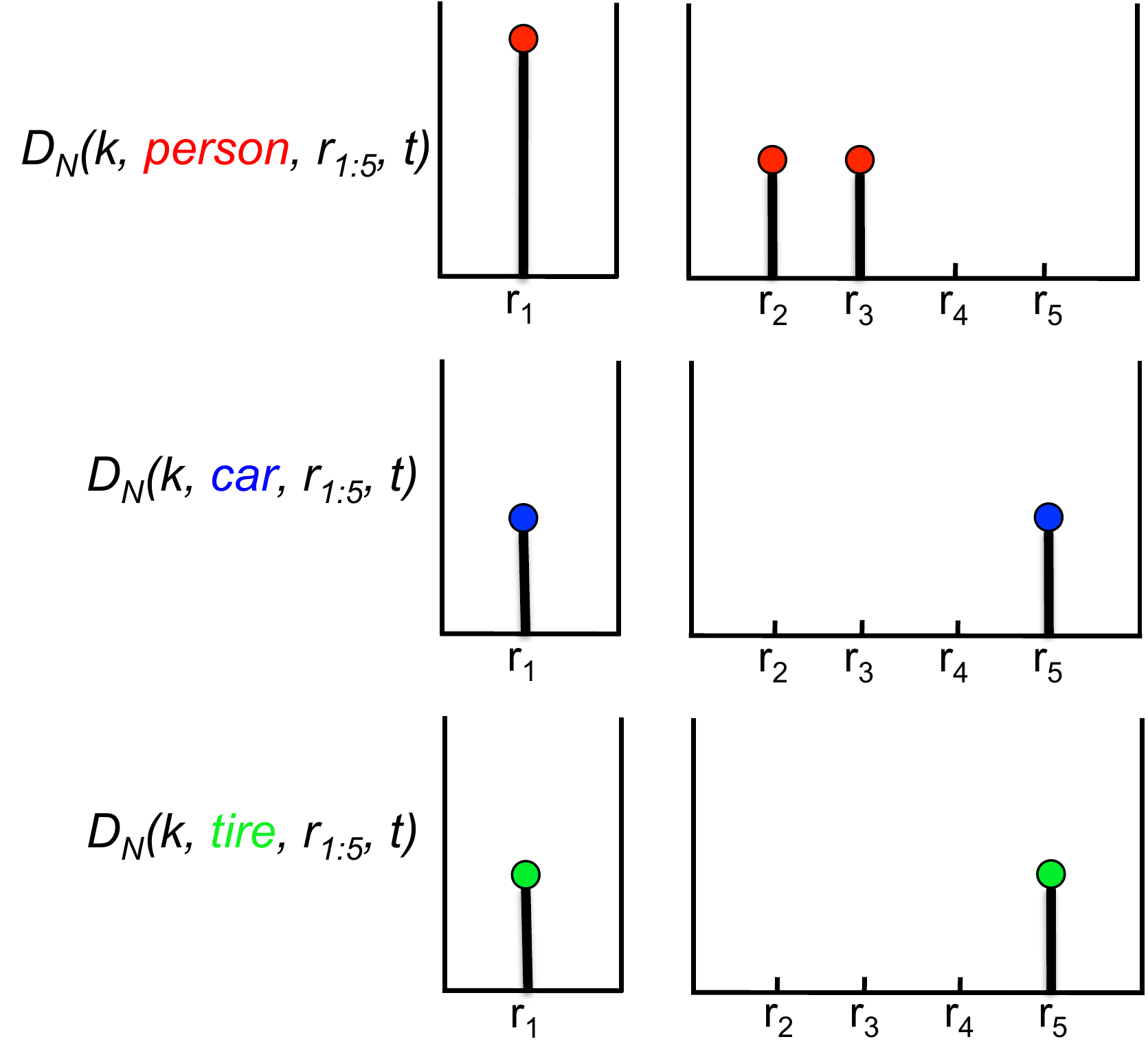}} 
\vspace{-0.15in}	
\caption{Left: Example detections on two successive keyframes with a subset of object detectors. Red, green and blue bounding boxes show person, tire, and car detections respectively. Orange and magenta dotted lines show a subset of the spatial pyramid. Right: Illustration of the detection count feature for the three categories on the first 5 grid cells of frame $k$. Grid cell $r_1$ has the extent of the whole image, $r_2, r_3, r_4, r_5$ are the cells in the orange quadrant clockwise from the top left corner. Best viewed in color.}
\vspace{-0.15in}
	\label{fig:main_figure}
\end{figure*}

There has been considerable progress in classifying videos using concept models formed from global feature representations. This approach learns a generic mapping from images to a set of predefined scene labels, which are then used as features to classify and/or retrieve a specific target query. Currently this class of method achieves state-of-the-art performance on scene-level image classification and many video retrieval tasks \cite{hauptmann, classemes}.

However, recognition and retrieval of events and activities may require discrimination based on specific objects in a scene; for example, the difference between a birthday party and a wedding ceremony may lie in particular clothing, objects such as candles and balloons in the scene, or even the type of cake.

Scene-level visual descriptors are generally inadequate to capture these fine-grained phenomena. Objects of interest may not cover the entire scene, and thus low-level visual descriptors that are pooled over the entire scene may fail to detect them. To be sensitive to the presence of individual objects in a scene requires visual analysis that considers windows or segments in a scene.

Recent advances in deformable part object detection have demonstrated that objects can be reliably found in many natural scenes, e.g., as demonstrated in the PASCAL VOC challenge \cite{PASCAL}. Progress is ongoing, but some categories are reasonably well detected. Most localized detection models employ either a top-down window scanning methodology, or a bottom-up segment generation strategy. Recently, the Object Bank model was proposed as a representation for indexing video based on the max pooled score map output of a bank of individual object detectors, each of which operated over windows of the scene \cite{objectbank}. The detectors are based on the deformable part model \cite{DPM} and were trained to recognize about 200 different object categories.

We extend this idea of using the output of object detectors to guide event recognition. While the Object Bank model provides a dense map of max pooled detection scores, it lacks a more immediate sense of whether or not there are objects present in the image and if so how many (e.g. the number of person detections presumably helps in differentiating ``Attempting a board trick'' from ``Flash mob gathering''). We propose to compute different statistics from the detection images from object detectors to capture this information more directly. The detection images are obtained by thresholding the score maps from the object detectors, applying non-maximum suppression, and pooling across all scales such that the detection images contains all the final detections.
Object Bank model omits all these steps that are standard in a detection pipeline which would result in a more sparse representation. We demonstrate that these steps along with the proposed detection count statistics lead to significantly better classification performance.
This image or key frame level representation is extended to a video level representation by mean and max pooling across keyframes which allows for an intuitive interpretation (see Section 3 for more detail). In light of the Object Bank model we call this new representation \textit{Detection Bank}.

We empirically show that combining features of this proposed video representation with existing global features significantly outperforms any of the methods alone on TRECVID MED 2011 data. 
Our results suggest that high-level visual features from a bank of object detections can be used to complement present multi modal concept detection systems (such as \cite{BBN}).

The rest of the paper is organized as follows. Section 2 gives
an overview of related work in image representations, object detection, and multimedia event detection. Section 3 describes the proposed Detection Bank representation in more detail. An evaluation of the approach is presented in Section 4 before Section 5 concludes the paper with a short discussion of results as well as future work ideas.

\section{Related Work}

The Spatial pyramid match (SPM) \cite{SPM} computes histograms of visual words from gradient orientation statistics at multiple scales. A generative topic model on bag of visual words to model the image scene generation process has been explored in \cite{feifei-scene}.
The Deformable part model (DPM) \cite{DPM} learns an object detector model from a weakly supervised setting where only the root location is supervised and the locations of parts are not known. It is the current state-of-the-art object detector on PASCAL VOC. 
Object Bank (OB) \cite{objectbank} uses the max pooled intermediate score maps from deformable part model object detectors as a global feature representation of images. In order to capture the presence of certain objects more directly in our representation we propose to explicitly build detection images and compute detection count statistics instead of only using the intermediate score maps.
Using DPM object detectors as high-level visual features for the task of multimedia event detection has been explored in \cite{BBN}. The authors report that their feature representation did not lead to a significant performance improvement (with the small exception of car detections and the ``Getting vehicle unstuck'' event). In contrast, we present an approach that leads to significant performance increase across all event categories.

\section{Detection Bank Representation}


As motivated earlier, object detections can be used to discriminate between certain events and activities, e.g., a large number of occurrences of flags and cars provides a strong cue to distinguish ``Parade'' from ``Flash mob gathering'' while they might both contain multiple persons. Similarly, detecting a large number of candles and balloons but no wedding dresses and bridegrooms speaks in favor of ``Birthday party'' instead of ``Wedding ceremony''.

As noted in the introduction, the Object Bank representation omits thresholding, non-maximum suppression, and pooling across different scales that are well-known mechanisms in the object detection community. Our Detection Bank representation explicitly builds detection images and computes the following detection count statistics (per object category) for each grid cell in a spatial pyramid (entire image, $2\times2$, and $4\times4$ grid): the sum of scores of detections within that cell (above a certain threshold), the number of detections, and a single bit that indicates whether or not there is a detection within in that cell. By mean and max pooling these statistics across keyframes of a video we obtain a meaningful video-level representation capturing, e.g., the maximum number of detections, the average number of detections, and an empirical estimate of the detection probability for each grid cell and object category. It will be demonstrated in Section 4 that these statistics contain discriminative information that is complementary to both scene-level features and max pooled detection score maps (as used in Object Bank).

\newcommand{\score}{\operatornamewithlimits{score}}

Formally, an object detector for category $c$ searches over every location $(x,y)$ of an image $\mathcal{I}$ and outputs $P$ predicted locations of the object in terms of bounding boxes $\mathcal{B}_{c}$ on $\mathcal{I}$,
$\mathcal{B}_{c} = \left[ \mathbf{b}_{c,1}, ~\hdots, ~ \mathbf{b}_{c,P} \right]$ where $\mathbf{b}_{c,i} = \left[  x_{i1}, y_ {i1}, x_{i2}, y_{i2},\score_i \right]$.
We consider the following three statistics per key frame $k$ from windowed object detectors at detection threshold $t$:
\begin{flalign}
&D_S \left( k, c, r, t \right) = \sum_{i=1}^P \mathbb{I} \left[ ~\overline{\mathbf{b}_{c,i}} \in \mathcal{I} \left( r \right) ~\right] \mathbb{I} \left[ s\left( \mathbf{b}_{c,i} \right) \geq t \right] s\left( \mathbf{b}_{c,i} \right) \nonumber&
\end{flalign} \vspace{-10pt}
\begin{flalign}
&D_N \left( k, c, r, t \right) = \sum_{i=1}^P \mathbb{I} \left[ ~\overline{\mathbf{b}_{c,i}} \in \mathcal{I} \left( r \right) ~\right] \mathbb{I} \left[ s\left( \mathbf{b}_{c,i} \right) \geq t \right]&
\end{flalign} \vspace{-10pt}
\begin{flalign}
&D_0 \left( k, c, r, t \right) = \mathbb{I} \left[ \sum_{i=1}^P \left( \mathbb{I} \left[ ~\overline{\mathbf{b}_{c,i}} \in \mathcal{I} \left( r \right) ~\right] \mathbb{I} \left[ s\left( \mathbf{b}_{c,i} \right) \geq t \right] \right)> 0 \right] \nonumber&
\end{flalign}
where $\mathbb{I}\left[ \cdot \right]$ is the indicator function, $t$ is a specific detection threshold, and $\overline{\mathbf{b}_{c,i}}$ denotes the center of the bounding box. $\mathcal{I} \left( r \right)$ denotes a spatial grid cell indexed by $r$. $s\left( \mathbf{b}_{c,i} \right)$ denotes the score of the bounding box. $D_S\left( k, c, r, t \right)$ is a detection statistic summing over the scores of detections within spatial grid region $r$ for a specific category $c$ and detection threshold $t$. $D_N\left( k, c, r, t \right)$ computes the number of detections within region $r$. $D_0\left( k, c, r, t \right)$ is a binary feature whether there are any detections within the region. 
As more conservative threshold values are applied the true positive rate increases at the expense of less detections.
Overall, thresholding scores and applying non-maximum suppression lead to a representation that is easier for a linear classifier to learn as demonstrated in Section 4.

Finally, we aggregate the key frame level statistics to the feature vector $\mathcal{F}$ for each video $\mathcal{V} = \{ {\mathcal{I}_1,~\hdots, ~\mathcal{I}_K} \}$ by mean and max pooling $\mathcal{P}$ across keyframes. Denote by $K,C,R,T$ the total number of frames, object categories, spatial regions, and threshold levels respectively.

\newcommand{\mean}{\operatornamewithlimits{mean}}

\begin{align}
\begin{split}
&\mathcal{P}\left(\mathcal{V}, c, r, t \right) : ~ \mathbb{R}^{ \displaystyle  3K} \mapsto ~\mathbb{R}^{ \displaystyle  3}\\
&\mathcal{P}_{max}\left(\mathcal{V}, c, r, t\right) = \begin{bmatrix} 
\displaystyle \max_{k \in K} D_S \left( k, c, r, t \right)\\
\displaystyle \max_{k \in K} D_N \left( k, c, r, t \right)\\
\displaystyle \max_{k \in K} D_0 \left( k, c, r, t \right)
\end{bmatrix} \nonumber\\
\end{split}
\end{align}
\begin{equation}
\mathcal{P}_{mean}\left(\mathcal{V}, c, r, t\right) = \begin{bmatrix} 
\displaystyle \mean_{k \in K} D_S \left( k, c, r, t \right)\\
\displaystyle \mean_{k \in K} D_N \left( k, c, r, t \right)\\
\displaystyle \mean_{k \in K} D_0 \left( k, c, r, t \right)\\
\end{bmatrix}
\end{equation}
\begin{flalign}
&\mathcal{F}_{max}\left( \mathcal{V} \right) = \left( \mathcal{P}_{max} \left( \mathcal{V}, c_1, r_1, t_1\right),\hdots,\mathcal{P}_{max} \left( \mathcal{V}, c_C, r_R, t_T\right)\right) \nonumber&
\end{flalign}
\begin{flalign}
&\mathcal{F}_{mean}\left( \mathcal{V} \right) = \left( \mathcal{P}_{mean} \left( \mathcal{V}, c_1, r_1, t_1\right),\hdots,\mathcal{P}_{mean} \left( \mathcal{V}, c_C, r_R, t_T\right)\right) \nonumber&
\end{flalign}

\noindent Figure ~\ref{fig:main_figure} illustrates our Detection Bank feature representation on two successive keyframes on a video from ``Changing a vehicle tire'' event.

\section{Experiments}

We evaluated the proposed detection-based video representation in the realm of multimedia event classification using a forced-choice classification paradigm on the TRECVID MED 2011 Event Kit that contains 2025 videos from 15 different events \cite{trecvid} (see Table \ref{table:events}). Note that this study differs from the detection paradigm on the larger DEV-T and DEV-O collections of the TRECVID MED 2011 data set in order to allow for an extensive comparison of the expressiveness and suitability of the proposed representation and study of the model parameters (e.g. number of models, number of thresholds, influence of different detection statistics etc.).

\begin{table}
\begin{center}
\scriptsize

	\begin{tabular}{|c|l|}
	\hline
	\textbf{EventID} & \textbf{Event Name}\\ 
	\hline
	1 & Attempting a board trick\\ 
	2 & Feeding an animal\\ 
	3 & Landing a fish\\ 
	4 & Wedding ceremony\\ 
	5 & Working on a woodworking project\\ 
	6 & Birthday party\\ 
	7 & Changing a vehicle tire\\ 
	8 & Flash mob gathering\\ 
	9 & Getting a vehicle unstuck\\ 
	10 & Grooming an animal\\ 
	11 & Making a sandwich\\ 
	12 & Parade\\ 
	13 & Parkour\\ 
	14 & Repairing an appliance\\ 
	15 & Working on a sewing project\\
	\hline
	\end{tabular}
\normalsize
\vspace{-0.12in} 
\caption{Short description for all the 15 events in the TRECVID MED 2011 dataset.}
\vspace{-0.33in} 
	\label{table:events}
\end{center}
\end{table}

\begin{figure}
\centering
\includegraphics[width=0.47\textwidth]{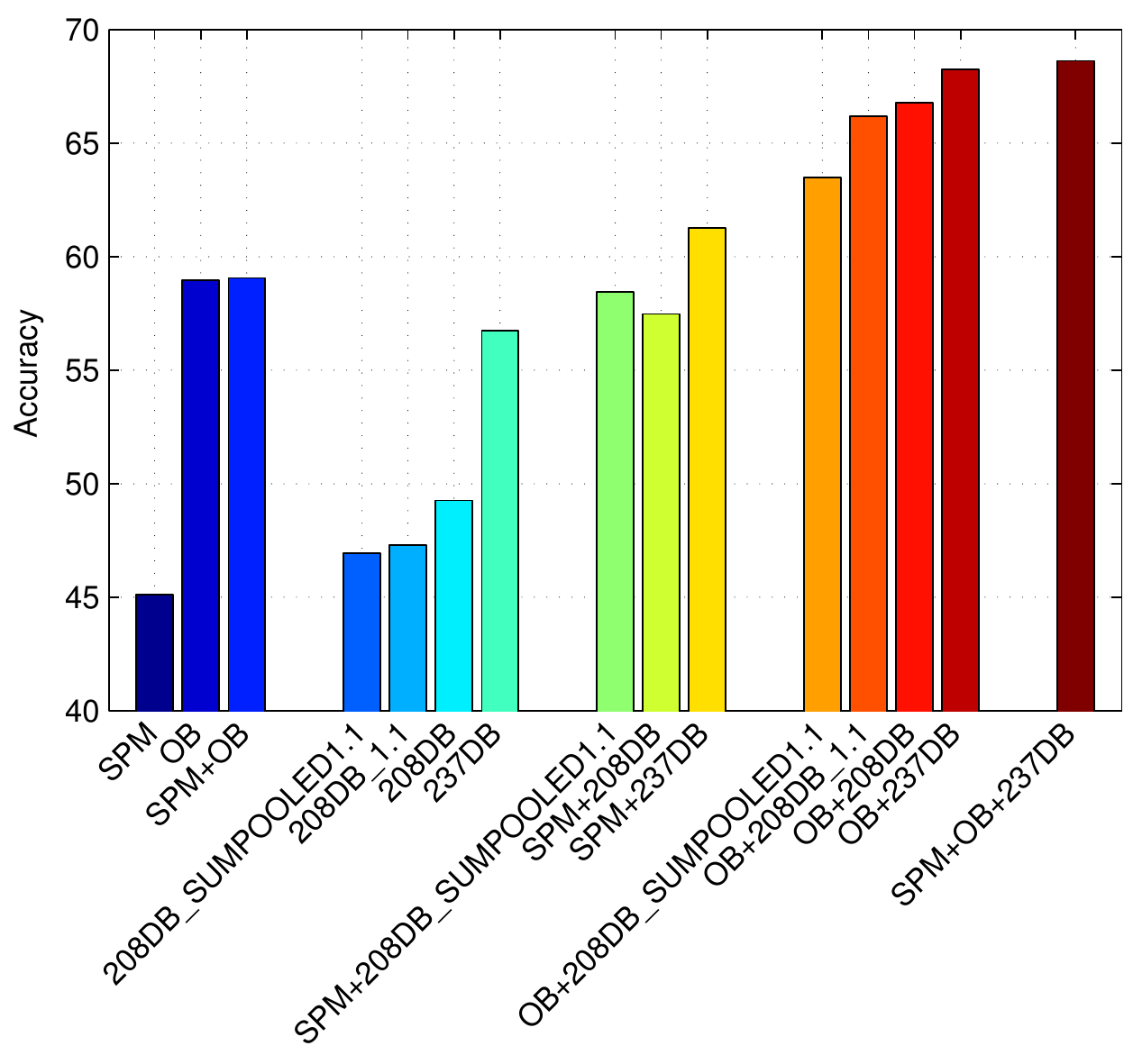} 
\vspace{-0.15in}	 
\caption{Classification accuracies on the TRECVID MED 2011 Event Kit for different feature combinations. The proposed Detection Bank (DB) representation provides complementary information to both scene-level (SPM) and window-based features (OB).}
\vspace{-0.17in} 
\label{fig:accuracybarplot}
\end{figure}

We randomly split the videos into a training (40\%), validation (20\%), and test (40\%) set. Support Vector Machines (SVM) are used to classify concatenated features of different representations as one of the 15 events.
We also experimented with Multiple Kernel Learning \cite{MKL} and Double Fusion \cite{doublefusion} which did not lead to a significant increase of performance in our case.
In our experience, linear kernel (one-vs-rest) SVM yielded the best performance in almost all cases. We also investigated both mean and max pooling techniques to pool the computed representation across all keyframes of a given video. Generally, max pooling worked best for Object Bank (OB) features while mean pooling worked best for both Spatial Pyramid Matching (SPM) features and Detection Bank (DB) features. Due to space constraints we only report the best classification accuracies per feature representation in Figure \ref{fig:accuracybarplot}. 

The first three bars represent the baseline accuracies using SPM (45.10\%) and OB (58.95\%) features (and their combination; 59.07\%). Object Bank performs already quite well due to the fact that it captures the likelihood of presence of certain objects that are relevant to the events (e.g. several animals, wheels, different tools, balloons, wedding gowns, bridegrooms etc.). 
The next four bars show the performance of our proposed representation alone where the number before DB refers to the number of DPM models used in the representation. We used 208 models from Object Bank, 20 models from PASCAL, and 9 additional models specifically trained for the defined task (boat, bread, cake, candle, fish, goat, jeep, scissors, and tire). These event-specific models were learned automatically by choosing categories based on available textual descriptions of the MED Events and acquiring available training data from ImageNet \cite{imagenet}.
Using only the $D_S$ statistic and only one threshold ($t=-1.1$) on all 208 models from Object Bank (208DB\_\-SUMPOOLED1.1) we obtain 46.94\%, using all statistics ($D_S, D_N, D_0$) but only one threshold (208DB\_1.1) 47.30\%, using all statistics and four thresholds ($t \in \{-1.1, -0.9, -0.7, -0.5\}$; 208DPM) 49.26\%, and using all 237 models, all statistics and four thresholds (237DB) 56.74\% accuracy. We have found these features to have only 30\% non-zero values (sparsity 70\%).

Our representation provides complementary discriminative information to both scene-level (SPM) and window-based object features (OB) by improved classification performance on the combined feature sets. Using only a relatively compact representation (SPM+\-208DB\_ SUMPOOLED1.1) we can obtain Object Bank-like performance of 58.46\% (8568D compared to 44604D). Using all three statistics and four thresholds (SPM+\-208DB), or additional models(SPM+237DB) we obtain 57.48\% and 61.27\%, respectively.

\begin{figure}
\hspace{0.15in}
\includegraphics[width=0.47\textwidth]{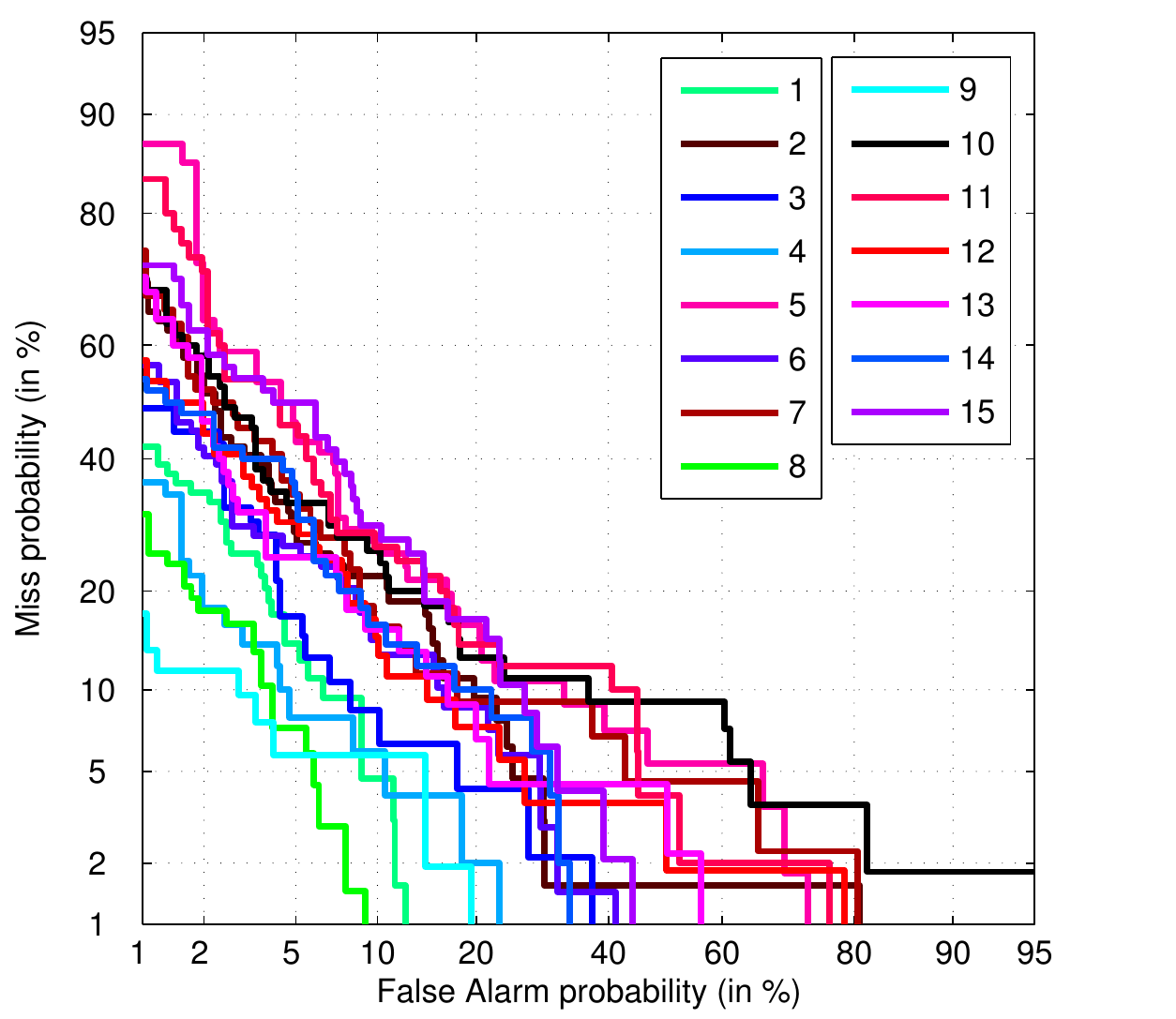} 
\vspace{-0.27in}	 
\caption{DET curves for all 15 events using the all features. Event IDs in order of intersection with horizontal axis are: 8, 1, 9, 4, 14, 3, 6, 15, 13, 5, 11, 12, 7, 2, and 10. The mapping from ID to Event is given in Table \ref{table:events}. Best viewed in color.}
\vspace{-0.17in} 
\label{fig:detcurve}
\end{figure}

The main result of this paper is that we can, without using more object models, use the proposed representation on top of Object Bank to significantly improve performance. Adding the $D_S$ statistic for 208 models with only one threshold we obtain 63.48\%, using all three statistics 66.18\%, using all statistics and four thresholds 66.79\%, and using the 29 additional models 68.26\% (almost 10\% higher than OB alone). If we also add SPM features to the last one, we reach the highest classification accuracy of 68.63\% (108528D feature).
Figure \ref{fig:detcurve} shows a DET curve for this feature combination for all 15 events.

In addition, we investigated the value of the additional  categories. Removing the 9 event-specific categories from the full combination of features leads to a drop of 0.7\% and removing all 29 additional categories leads to a drop of 1.2\% in classification accuracy.

\vspace{-0.06in}
\section{Conclusions}
\vspace{-0.01in}
We proposed a feature representation on videos which we call Detection Bank based on the detection images from a large number of windowed object detectors. The Detection Bank representation provides complementary discriminative information to current state-of-the-art image representations such as Spatial Pyramid Matching and Object Bank. We demonstrated that these combined with our Detection Bank representation provide a significant improvement in multimedia event classification on TRECVID MED 2011 data.

As future work, we plan to evaluate the proposed representation in a detection scenario on the full TRECVID MED 2011 data set. Further, we want to investigate the influence of event-specific object categories that could be learned on the fly and to what degree detecting meaningful categories differs from simply having a large number of independent measurements from random filter responses.

%
\vspace{-0.1in}
\bibliographystyle{abbrv}
\scriptsize
\bibliography{acmmm}

\vspace{.05in}
\scriptsize 
\noindent ACKNOWLEDGMENT: Supported by the Intelligence Advanced Research Projects Activity (IARPA) via Department of Interior National Business Center contract number D11PC20066. The U.S. Government is authorized to reproduce and distribute reprints for Governmental purposes notwithstanding any copyright annotation thereon.
Disclaimer: The views and conclusion contained herein are those of the authors and should not be interpreted as necessarily representing the official policies or endorsement, either expressed or implied, of IARPA, DOI/NBC, or the U.S. Government. \vspace{-0.15in}

\end{document}